\documentclass[superscriptaddress,amsmath,amssymb,aps,pre]{revtex4-2}
\usepackage{bm}
\usepackage{lineno}
\usepackage{graphicx}
\usepackage{xcolor}
\usepackage[margin=0.9in]{geometry}
\usepackage[bookmarks = false, pdfstartview = {FitB}, colorlinks = true, allcolors = blue]{hyperref}

\makeatletter
\renewcommand{\@fnsymbol}[1]{\ensuremath{\ifcase#1\or \dagger\or *\or \ddagger\or
   \mathsection\or \mathparagraph\or \|\or **\or \dagger\dagger
   \or \ddagger\ddagger \else\@ctrerr\fi}}
\makeatother

\begin{document}

\title{Mixing in chaotic flows with swimming bacteria} 
\author{Ranjiangshang Ran}
\thanks{ranr@seas.upenn.edu}
\author{Quentin Brosseau}
\affiliation{Department of Mechanical Engineering and Applied Mechanics, University of Pennsylvania, Philadelphia, PA 19104, USA}
\author{Brendan C. Blackwell}
\affiliation{Department of Mechanical Engineering and Applied Mechanics, University of Pennsylvania, Philadelphia, PA 19104, USA}
\affiliation{Department of Physics and Astronomy, Northwestern University, Evanston, IL 60208, USA}
\author{Boyang Qin}
\affiliation{Department of Mechanical Engineering and Applied Mechanics, University of Pennsylvania, Philadelphia, PA 19104, USA}
\affiliation{Department of Mechanical and Aerospace Engineering, Princeton University, Princeton, NJ 08544, USA}
\author{Rebecca L. Winter}
\author{Paulo E. Arratia}
\thanks{parratia@seas.upenn.edu}
\affiliation{Department of Mechanical Engineering and Applied Mechanics, University of Pennsylvania, Philadelphia, PA 19104, USA}

\date{\today}

\begin{abstract}
This is a manuscript accepted for publication on \textit{Physical Review Fluids}, Gallery of Fluid Motion special issue. The manuscript is associated with a poster winner of the 39th Annual Gallery of Fluid Motion Award, for work presented at the 74th Annual Meeting of the American Physical Society's Division of Fluid Dynamics (Phoenix, AZ, USA 2021).
\end{abstract}

\maketitle 

\begin{figure}[b!]\label{fig1}
\centering
\includegraphics[width=3.37in]{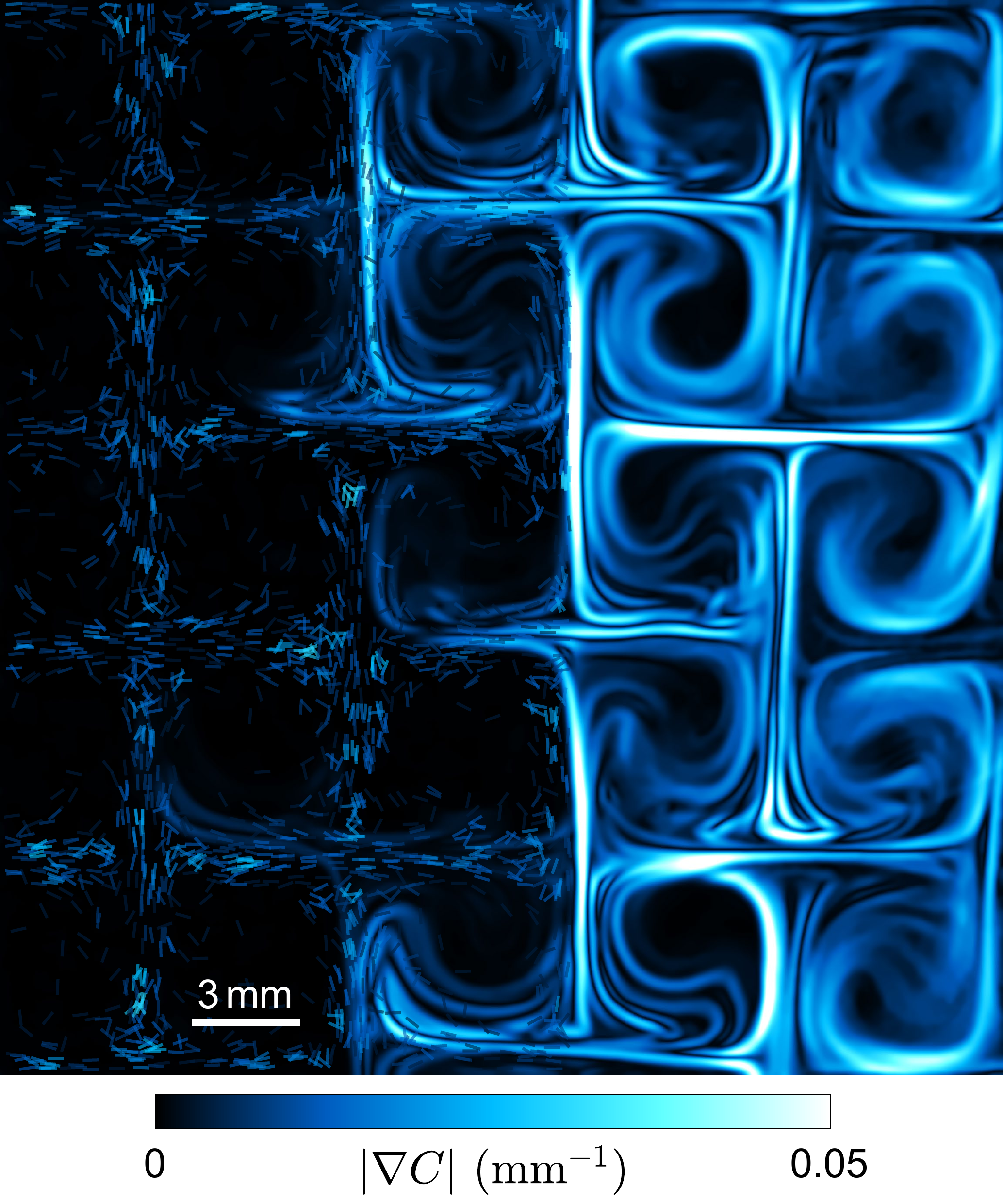}
\vspace{-2ex}
\caption{Intriguing dye mixing pattern produced by chaotic advection, visualized by the dye concentration gradient field, $\vert\nabla C\vert$. Also plotted on the left is the spatial distribution of the swimming bacteria; strong accumulations of bacteria are found near dynamical flow structures. These flow structures coincide with the exact same structures visualized by dye gradient patterns. The size of the imaged region is $28~\mathrm{mm}\times30~$mm. The original image is available online at: \href{https://doi.org/10.1103/APS.DFD.2021.GFM.P0036}{https://doi.org/10.1103/APS.DFD.2021.GFM.P0036}.}
\end{figure}

How does the swimming motion of microorganisms affect the transport and mixing properties of a passive scalar in complex flows? The answers to this question can potentially improve our understanding of the spread of pollutants (e.g. algal blooms, oil spills) in oceans and lakes, as well as stimulate the development of useful applications in biofuel and vaccine productions. Fluid flows can affect the swimming behaviors of microorganisms, causing them to form aggregates, clusters and patches near flow structures \cite{Stocker_AR_2019}; these aggregates or clusters, in turn, can change the local fluid properties and alter the fluid flows \cite{Ran_PNAS_2021}. Such flow-microorganism interaction gives rise to unique transport and mixing phenomena of active swimming particles that are not observed for passive particles.

In our experiments, we investigate the effects of activity on the chaotic mixing of a passive scalar in dilute suspensions of \textit{Escherichia coli}. Figure \href{fig2}{2}(a) shows the experimental setup, where a thin layer of active suspension is placed above an array of magnets of alternating polarity. As a sinusoidal voltage is imposed, the induced Lorentz force creates a cellular flow with an array of vortices, the flow field of which is shown in Fig. 2(b). Half of the fluid layer is initially labelled by a fluorescent dye (the passive scalar). The dye is illuminated under black light, and recorded by a high-resolution camera (Flare 4M180) operating at 5 frames per second and $2000^2$-pixel resolution. A UV-light filter (Tiffen, Yellow 12) is used such that the light intensity is accurately proportional to the dye concentration field, $C$. More details on the experimental setup can be found in \cite{Ran_PNAS_2021}. As the flow is turned on, the labelled fluid progressively penetrates to the unlabelled portion and produces complex dye mixing patterns.

\begin{figure}[t!]\label{fig2}
\centering
\includegraphics[width=6.1in]{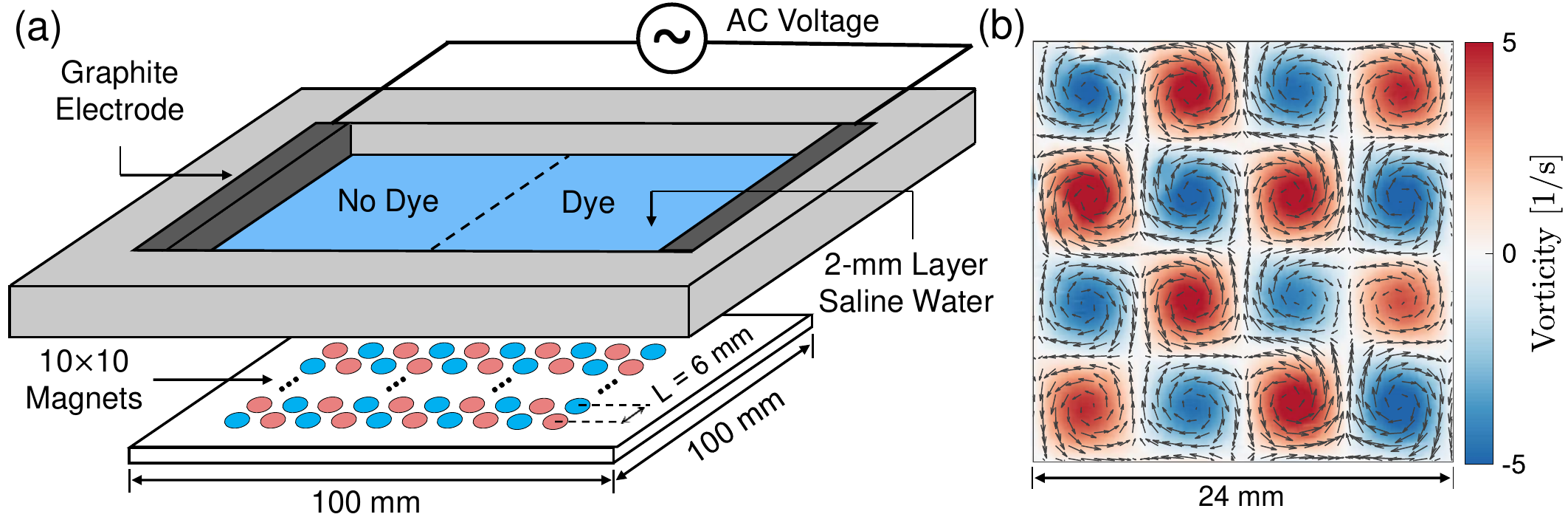}
\vspace{-2ex}
\caption{(a) Schematic of the flow cell apparatus. A thin layer of saline water containing swimming \textit{Escherichia coli} is placed above an array of magnets of alternating polarity, illustrated by different colors. As a sinusoidal voltage is imposed through the
electrodes, the induced Lorentz force of the migrating ions in the fluid layer creates chaotic mixing. The right half of the fluid layer is initially labeled with fluorescent dye. (b) Vorticity field (colormap) and velocity field (arrows) of the mixing flow generated by the flow cell. The imaged area is a $4\times4$ portion of the $10\times10$ vortex array, with a size of $24~\mathrm{mm}\times24~$mm.}
\end{figure}

Figure \href{fig1}{1} shows an intriguing dye mixing pattern produced by the flow. The pattern is calculated from the gradient magnitude of the dye concentration field, $\vert\nabla C\vert$, which characterizes the regions of high passive scalar transport and coincide with the dynamical structures of the flow \cite{Voth_PRL_2002}. These flow structures, known as Lagrangian coherent structures (LCSs), are regions of locally extremal flow strain that maximally attract or repel nearby fluid particle trajectories \cite{Haller_LCS_2013}. It has been recently shown that active particles can interact with these LCSs \cite{Khurana_PoF_2012}. Here, we demonstrate such interactions by showing bacteria accumulation along the exact same flow structures that are also visualized by the concentration gradient field (Fig. \href{fig1}{1}, left). We further show that such accumulation near the dynamical flow structures have nontrivial effects on the transport and mixing of the contaminant, both locally and globally; see details in \cite{Ran_PNAS_2021}. 

\vspace{+2ex}
This work was supported by the US National Science Foundation (NSF) through Grant No. DMR-1709763.


\bibliography{reference.bib}

\end{document}